\documentclass[a4paper]{report}
\usepackage[utf8]{inputenc}
\usepackage[T1]{fontenc}
\usepackage{RJournal}
\usepackage{amsmath,amssymb,array}
\usepackage{booktabs}
\usepackage{bm}

\def\bbeta{\bm \beta}

\def \ep{\epsilon}
\def \mmin{\wedge}

\begin{document}

\sectionhead{Contributed research article}
\volume{XX}
\volnumber{YY}
\year{20ZZ}
\month{AAAA}

\begin{article}
\title{A Fast and Scalable Implementation Method for Competing Risks Data with the R Package \CRANpkg{fastcmprsk}}
\author{by Eric S. Kawaguchi, Jenny I. Shen, Gang Li, and Marc A. Suchard}

\maketitle

\abstract{
Advancements in medical informatics tools and high-throughput biological experimentation make large-scale biomedical data routinely accessible to researchers. Competing risks data are typical in biomedical studies where individuals are at risk to more than one cause (type of event) which can preclude the others from happening. The \cite{fine1999proportional} proportional subdistribution hazards model is a popular and well-appreciated model for competing risks data and is currently implemented in a number of statistical software packages. However, current implementations are not computationally scalable for large-scale competing risks data. We have developed an R package, \CRANpkg{fastcmprsk}, that uses a novel forward-backward scan algorithm to significantly reduce the computational complexity for parameter estimation by exploiting the structure of the subject-specific risk sets. Numerical studies compare the speed and scalability of our implementation to current methods for unpenalized and penalized Fine-Gray regression and show impressive gains in computational efficiency.}

\section{Introduction} 

\label{sec3:intro}
Competing risks time-to-event data arise frequently in biomedical research when subjects are at risk for more than one type of possibly correlated events or causes and the occurrence of one event precludes the others from happening. 
For example, one may wish to study time until first kidney transplant for kidney dialysis patients with end stage renal disease. Terminating events such as death, renal function recovery, or discontinuation of dialysis are considered competing risks as their occurrence will prevent subjects from receiving a transplant.  When modeling competing risks data the cumulative incidence function (CIF), the probability of observing a certain cause while taking the competing risks into account, is oftentimes a quantity of interest. 

The most commonly-used model to draw inference about the covariate effect on the CIF and to predict the CIF dependent on a set of covariates is the Fine-Gray proportional subdistribution hazards model \citep{fine1999proportional}. Various statistical packages for estimating the parameters of the Fine-Gray model are popular within the {R} programming language. One package, among others, is the \CRANpkg{cmprsk} package. The \CRANpkg{riskRegression} package, initially implemented for predicting absolute risks \citep{gerds2012absolute}, uses a wrapper that calls the \CRANpkg{cmprsk} package to perform Fine-Gray regression. \cite{scheike2011analyzing} provide \CRANpkg{timereg} that allows for general modeling of the cumulative incidence function and includes the Fine-Gray model as a special case. The \CRANpkg{survival} package also performs Fine-Gray regression but does so using a weighted Cox \citep{cox1972regression} model. Over the past decade, there have been several extensions to the Fine-Gray method that also result in useful packages. The \CRANpkg{crrSC} package allows for the modeling of both stratified \citep{zhou2011competing} and clustered \citep{zhou2012competing} competing risks data. \cite{kuk2013model} propose a stepwise Fine-Gray selection procedure and develop the \CRANpkg{crrstep} package for implementation. \cite{fu2017penalized} then introduce penalized Fine-Gray regression with the corresponding \CRANpkg{crrp} package. 

A contributing factor to the computational complexity for general Fine-Gray regression implementation is parameter estimation. Generally, one needs to compute the log-pseudo likelihood and its first and second derivatives with respect to its regression parameters for optimization. Calculating these quantities is typically of order $O(n^2)$, where $n$ is the number of observations in the dataset, due to the repeated calculation of the subject-specific risk sets. With current technological advancements making large-scale data from  electronic health record (EHR) data systems routinely accessible to researchers, these implementations quickly become inoperable or grind-to-a-halt in this domain. 
For example, \cite{kawaguchi2019scalable} reported a runtime of about 24 hours to fit a LASSO
regularized Fine-Gray regression
 on a subset of the United States Renal Data Systems (USRDS) with $n =125, 000$
subjects using an existing R package
\CRANpkg{crrp}.
To this end, we note that for time-to-event data with no competing risks, \cite{simon2011regularization}, \cite{breheny2011coordinate}, and \cite{ mittal2013high}, among many others, have made significant progress in reducing the computational complexity for the \cite{cox1972regression} proportional hazards model from $O(n^2)$ to $O(n)$ by taking advantage of the cumulative structure of the risk set. However, the counterfactual construction of the risk set for the Fine-Gray model does not retain the same structure and presents a barrier to reducing the complexity of the risk set calculation. To the best of our knowledge, no further advancements in reducing the computational complexity required for calculating the subject-specific risk sets exists. 

The contribution of this work is the development of an {R} package \CRANpkg{fastcmprsk} which implements a novel forward-backward scan algorithm \citep{kawaguchi2019scalable} for the Fine-Gray model. By taking advantage of the ordering of the data and the structure of the risk set, we can calculate the log-pseudo likelihood and its derivatives, which are necessary for parameters estimation, in $O(n)$ calculations rather than $O(n^2)$. As a consequence, our approach is scalable to large competing risks datasets and outperforms competing algorithms for both penalized and unpenalized parameter estimation. 

The paper is organized as follows. In the next section, we briefly review the basic definition of the Fine-Gray proportional subdistribution hazards model, the CIF, and penalized Fine-Gray regression. We highlight the computational challenge of lineaizing estimation for the Fine-Gray model and introduce the forward-backward scan algorithm of \cite{kawaguchi2019scalable} in Section \ref{s3:scan}. Then, in Section \ref{s3:pkg} we describe the main functionalities of the \CRANpkg{fastcmprsk} package that we developed for {R} which utilizes the aforementioned algorithm for unpenalized and penalized parameter estimation and CIF estimation. We perform simulation studies in Section \ref{s3:sim} to compare the performance of our proposed method to some of their popular competitors. The \CRANpkg{fastcmprsk} package is readily available on the Comprehensive R Archive Network (CRAN) at  \url{https://CRAN.R-project.org/package=fastcmprsk}.


\section{Preliminaries}
\subsection{Data structure and model}
\label{s3:estimator}
We first establish some notation and the formal definition of the data generating process for competing risks. 
For subject $i = 1, \ldots, n$, let $T_i$, $C_i$, and $\ep_i$ be the event time, possible right-censoring time, and cause (event type), respectively. Without loss of generality assume there are two event types $\ep \in \{1, 2\}$ where $\ep = 1$ is the event of interest (or primary event) and $\ep = 2$ is the competing risk. With the presence of right-censoring, we generally observe $X_i = T_i \mmin C_i$, $\delta_i = I(T_i \leq C_i)$, where $a \mmin b = \min(a, b)$ and $I(\cdot)$ is the indicator function. Letting $\mathbf{z}_i$ be a $p$-dimensional vector of time-independent subject-specific covariates, competing risks data consist of the following independent and identically distributed quadruplets $\{(X_i, \delta_i, \delta_i \ep_i, \mathbf{z}_i)\}_{i=1}^n$. Assume that there also exists a $\tau$ such that  1) for some arbitrary time $t$, $t \in [0, \tau]$ ; 2) $\Pr(T_i > \tau) > 0$ and $\Pr(C_i > \tau) >0$  for all $i = 1,\ldots, n$, and that for simplicity, no ties are observed.

The CIF for the primary event  conditional on the covariates $\mathbf{z} = (z_1, \ldots, z_p)$ is $F_1(t; \mathbf{z}) = \Pr(T \leq t, \epsilon = 1|\mathbf{z})$.  To model the covariate effects on $F_1(t; \mathbf{z})$, \cite{fine1999proportional} introduced the now well-appreciated proportional subdistribution hazards (PSH) model: 
\begin{align}
\label{eq3:pshmodel}
h_1(t| \mathbf{z}) = h_{10}(t) \exp(\mathbf{z}^\prime\bbeta),
\end{align}
where \begin{align*}
h_1(t| \mathbf{z}) & =  \lim_{\Delta t \to 0} \frac{\Pr\{t \leq T \leq t + \Delta t, \epsilon = 1 | T \geq t \cup (T \leq t \cap \epsilon \neq 1), \mathbf{z}\}}{\Delta t} 
\\&  = -\frac{d}{dt} \log\{1 - F_1(t; \mathbf{z})\}
\end{align*}
is a subdistribution hazard \citep{gray1988class}, 
$h_{10}(t)$ is a completely unspecified baseline subdistribution hazard, and $\bbeta$ is a $p \times 1$ vector of regression coefficients. 
As \cite{fine1999proportional} mentioned, the risk set associated with $h_1(t; \mathbf{z})$ is somewhat 
unnatural as it includes subjects who are still at risk $(T \geq t)$ and those who have already observed the competing risk prior to time $t$ ($T \leq t \cap \epsilon \neq 1$). However, this construction is useful  for direct modeling of the CIF.

\subsection{Parameter estimation for unpenalized Fine-Gray regression}
Parameter estimation and large-sample inference of the PSH model follows from the log-pseudo likelihood:
\begin{align}
\label{eq3:lpp}
l(\bbeta) = \sum_{i=1}^n  \int_0^\infty \left[ \mathbf{z}_i^\prime \bbeta - \ln \left\{ \sum_k \hat{w}_k(u) Y_k(u) \exp \left(\mathbf{z}_k^\prime \bbeta \right) \right\} \right] \hat{w}_i(u)dN_i(u),
\end{align}
where $N_i(t) = I(X_i \leq t, \ep_i = 1)$, $Y_i(t) = 1 - N_i(t-)$, and $\hat{w}_i(t)$ is a time-dependent weight based on the inverse probability of censoring weighting (IPCW) technique \citep{robins1992recovery}. To parallel \cite{fine1999proportional}, we define the  IPCW for subject $i$ at time $t$ as $\hat{w}_i(t) = I(C_i \geq T_i \mmin t)\hat{G}(t)/\hat{G}(X_i \mmin t)$, where $G(t) = \Pr(C \geq t)$ is the survival function of the censoring variable $C$ and $\hat{G}(t)$ is the Kaplan-Meier estimate for $G(t)$. However, we can generalize the IPCW to allow for dependence between $C$ and $\mathbf{z}$.

Let $\hat{\bbeta}_{mple} = \arg \min_{\bbeta} \{-l(\bbeta)\}$ be the maximum pseudo likelihood estimator of $\bbeta$. \cite{fine1999proportional} investigate the large-sample properties of $\hat{\bbeta}_{mple}$ and prove that, under certain regularity conditions, 
\begin{align}
\label{eq3:asymptotic}
\sqrt{n}(\hat{\bbeta}_{mple} - \bbeta_0) \to N(0, \Omega^{-1} \Sigma \Omega^{-1}),
\end{align}
where $\bbeta_0$ is the true value of $\bbeta$, $\Omega$ is the limit of the negative of the partial derivative matrix of the score function evaluated at $\bbeta_0$, and $\Sigma$ is the variance-covariance matrix of the limiting distribution of the score function. {We refer readers to \cite{fine1999proportional} for more details on $\Omega$ and $\Sigma$}. The package \CRANpkg{cmprsk} implements this variance estimation procedure. 

\subsection{Estimating the cumulative incidence function}
An alternative interpretation of the coefficients from the Fine-Gray model is to model their effect on the CIF. Using a Breslow-type estimator \citep{breslow1974covariance}, we can obtain a consistent estimate for $H_{10}(t) = \int_0^t h_{10}(s)ds$ through
\begin{align*}
\hat{H}_{10}(t) = \frac{1}{n} \sum_{i=1}^n \int_0^t \frac{1}{\hat{S}^{(0)}(\hat{\bbeta}, u)}\hat{w}_i(u)dN_i(u),
\end{align*}
where $\hat{S}^{(0)}(\hat{\bbeta}, u) = n^{-1} \sum_{i=1}^n \hat{w}_i(u)Y_i(u)\exp(\mathbf{z}_i^\prime\hat{\bbeta})$.
The predicted CIF, conditional on $\mathbf{z} = \mathbf{z}_0$, is then
\begin{align*}
\hat{F}_1(t;\mathbf{z}_0) = 1 - \exp\left\{\int_0^t \exp(\mathbf{z}^\prime_0 \hat{\bbeta})d\hat{H}_{10}(u)\right\}.
\end{align*}
We refer the readers to Appendix B of \cite{fine1999proportional} for the large-sample properties of $\hat{F}_1(t; \mathbf{z}_0)$. The quantities needed to estimate $\int_0^t d\hat{H}_{10}(u)$ are already precomputed when estimating $\hat{\bbeta}$. \cite{fine1999proportional} proposed a resampling approach to calculate confidence intervals and confidence bands for $\hat{F}_1(t; \mathbf{z}_0)$.

\subsection{Penalized Fine-Gray regression for variable selection}
\label{s3:pen}

Oftentimes, reserachers are interested in identifying which covariates have an effect on the CIF. Penalization methods  \citep{tibshirani1996regression, fan2001variable, zou2006adaptive, zhang2010regularization} offer a popular way to perform variable selection and parameter estimation simultaneously through minimizing the objective function
\begin{align}
Q(\bbeta) = -l(\bbeta) + \sum_{j=1}^p p_\lambda(|\beta_j|),
\end{align}
where $l(\bbeta)$ is defined in (\ref{eq3:lpp}), $p_{\lambda}(|\beta_j|)$ is a penalty function where the sparsity of the model is controlled by the non-negative tuning parameter $\lambda$. \cite{fu2017penalized} recently extend several popular variable selection procedures - LASSO \citep{tibshirani1996regression}, SCAD \citep{fan2001variable}, adaptive LASSO \citep{zou2006adaptive}, and MCP \citep{zhang2010nearly} - to the Fine-Gray model, explore its asymptotic properties under fixed model dimension, and develop the {R} package \CRANpkg{crrp} \citep{crrp} for implementation. Parameter estimation in the \CRANpkg{crrp} package employs a cyclic coordinate algorithm.

The sparsity of the model depends heavily on the choice of the tuning parameters. Practically, finding a suitable (or optimal) tuning parameter involves applying a penalization method over a sequence of possible candidate values of $\lambda$ and finding the $\lambda$ that minimizes some metric such as the Bayesian information criterion \citep{schwarz1978estimating} or generalized cross validation measure \citep{craven1978smoothing}. A more thorough discussion on tuning parameter selection can partially be found in 
 \cite{wang2007tuning, zhang2010regularization, wang2011consistent,  fan2013tuning, fu2017penalized, ni2018tuning}.

\section{Parameter estimation in linear time}
\label{s3:scan}
{ Whether interest is in fitting an unpenalized model or a series of penalized models used for variable selection, one will need to minimize the negated log-pseudo (or penalized log-pseudo) likelihood. While current implementations can readily fit small to moderately-sized datasets, where the sample size can be in the hundreds to thousands, we notice that these packages grind to a halt for large-scale data such as, electronic health records (EHR) data or cancer registry data, where the number of observations easily exceed tens of thousands, as illustrated later in  Section \ref{s3:crr} (Table 2) on some simulated large competing risks data.
The primary computational bottleneck for estimation the parameters of the Fine-Gray model is due to the calculation of the log-pseudo likelihood and its derivatives, which are required for commonly-used optimization routines. For example, the cyclic coordinate descent algorithm requires the score function
 \begin{align}
\label{eq3:psh_score}
\dot{l}_j(\bbeta) = & \sum_{i=1}^n I(\delta_i\ep_i = 1) z_{ij} - \sum_{i = 1}^n I(\delta_i\ep_i = 1) \frac{ \sum_{k \in R_i} z_{kj} \tilde{w}_{ik}  \exp(\eta_k)}{\sum_{k \in R_i} \tilde{w}_{ik} \exp(\eta_k)}, 
\end{align}
and the Hessian diagonals
\begin{align}
\label{eq3:psh_hess}
\ddot{l}_{jj}(\bbeta) = & \sum_{i=1}^n I(\delta_i\ep_i = 1) \left[ \frac{ \sum_{k \in R_i} z_{kj}^2 \tilde{w}_{ik}  \exp(\eta_k)}{\sum_{k \in R_i} \tilde{w}_{ik} \exp(\eta_k)} -  \left\{ \frac{ \sum_{k \in R_i} z_{kj} \tilde{w}_{ik}  \exp(\eta_k)}{\sum_{k \in R_i} \tilde{w}_{ik} \exp(\eta_k)} \right\}^2 \right], 
\end{align}
where  $$\tilde{w}_{ik} = \hat{w}_k(X_i) = \hat{G}(X_i) / \hat{G}(X_i \mmin X_k), \quad k \in R_i,$$ 
$R_i = \{y:(X_y \geq X_i) \cup (X_y \leq X_i \cap \epsilon_y =  2)\}$ and
$\eta_k = \mathbf{z}_k^\prime\bbeta$ for optimization. While the algorithm itself is quite efficient, especially for estimating sparse coefficients, direct evaluation of (\ref{eq3:psh_score}) and (\ref{eq3:psh_hess}) will require $O(n^2)$ operations since for each $i$ such that $\delta_i \epsilon_i = 1$ we must identify all $y \in \{1, \ldots, n\}$ such that either $X_y \geq X_i$ or $(X_y \leq X_i \cap \epsilon_y = 2)$. As a consequence, parameter estimation will be computationally taxing for large-scale data since runtime will scale quadratically with $n$. We verify this in Section \ref{s3:sim} for the \CRANpkg{cmprsk} and \CRANpkg{crrp} packages. To the best of our knowledge, prior to \cite{kawaguchi2019scalable}, previous work on reducing the computational of parameter estimation from $O(n^2)$ to a lower order has not been developed. 

Before moving forward we will first consider the Cox proportional hazards model for right-censored data, which can be viewed as a special case of the Fine-Gray model when competing risks are not present (i.e. $R_i = \{y: X_y \geq X_i\}$, $\tilde{w}_{ik} = 1$ for all $k \in R_i$, $\epsilon_i = 1$ whenever $\delta_i = 1$). Again, direct calculation of quantities such as the log-partial likelihood and score function will still require $O(n^2)$ computations; however, one can show that when event times are arranged in decreasing order, the risk set is monotonically increasing as a series of cumulative sums. Once we arrange the event times in decreasing order, these quantities can be calculated in $O(n)$ calculations. The simplicity of the data manipulation and implementation makes this approach widely adopted in several R packages for right-censored data including the \CRANpkg{survival}, \CRANpkg{glmnet}, \CRANpkg{ncvreg}, and \CRANpkg{Cyclops} packages.  

Unfortunately, the risk set associated with the Fine-Gray model does not retain the same cumulative structure. \cite{kawaguchi2019scalable} propose a novel forward-backward scan algorithm that reduces the computational complexity associated with parameter estimation from $O(pn^2)$ to $O(pn)$, allowing for the analysis of large-scale competing risks data in linear time. Briefly, the risk set $R_i$ partitions into two disjoint subsets: $R_i(1) = \{y:X_y \geq X_i\}$ and $R_i(2) = \{y:(X_y \leq X_i \cap \epsilon_y =  2) \}$, were $R_i(1)$ is the set of observations that have an observed event time after $X_i$ and $R_i(2)$ is the set of observations that have observed the competing event before time $X_i$. Since $R_i(1)$ and $R_i(2)$ are disjoint, the summation over $k \in R_i$ can be written as two separate summations, one over $R_i(1)$ and one over $R_i(2)$. The authors continue to show that the summation over $R_i(1)$ is a series of cumulative sums as the event times decrease while the summation over $R_i(2)$ is a series of cumulative sums as the event times increase. Therefore, by cleverly separating the calculation of both summations, (\ref{eq3:psh_score}), (\ref{eq3:psh_hess}), and consequently (\ref{eq3:lpp}) are available in $O(n)$ calculations. We will show the computational advantage of this approach for parameter estimation over competing R packages in Section \ref{s3:sim}.

}
\section{The fastcmprsk package}
\label{s3:pkg}
We utilize this forward-backward scan algorithm of \cite{kawaguchi2019scalable} for both penalized and unpenalized parameter estimation for the  Fine-Gray model in linear time. Furthermore, we also develop scalable methods to estimate the predicted CIF and its corresponding confidence interval/band. For convenience to researchers and readers, a function to simulate two-cause competing risks data is also included. Table 1 provides a summary of the currently available functions provided in \CRANpkg{fastcmprsk}. We briefly detail the use of { some of the key} functions below.

\begin{table}[t]
\centering
\begin{tabular}{ll}
  \toprule
  Function name & Basic description \\
  \hline
  \em{Modeling functions} & \\
  \code{fastCrr}  & Fits unpenalized Fine-Gray regression \\
  \code{fastCrrp} & Fits penalized Fine-Gray regression\\
  \midrule
  \em{Utilities} & \\
  \code{Crisk} & Creates a competing risk object to be used as the\\
  & response variable for \code{fastCrr} and \code{fastCrrp}\\
  \code{varianceControl } & Options for bootstrap variance for \code{fastCrr}.\\
  \code{simulateTwoCauseFineGrayModel} & Simulates two-cause competing risks data\\
  \midrule
  {\em{S3 methods for fastCrr}} & \\
  \code{AIC} & Generic function for calculating AIC \\
  \code{coef} & Extracts model coefficients \\
  \code{confint} & Computes confidence intervals for \\
  & parameters in the model\\
  \code{logLik} & Extracts the model log-pseudo likelihood \\
  \code{predict} & Predict the cumulative incidence function \\
  & given \code{newdata} using model coefficients. \\
  \code{summary} & Print ANOVA table \\
  \code{vcov} & Returns bootstrapped variance-covariance matrix \\
  & if \code{variance = TRUE}. \\
    \midrule
  {\em{S3 methods for fastCrrp}} & \\
     \code{AIC} & Generic function for calculating AIC \\
  \code{coef} & Extracts model coefficients \\
  & for each tuning parameter $\lambda$. \\
  \code{logLik} & Extracts the model log-pseudo likelihood  \\
  & for each tuning parameter $\lambda$. \\
  \code{plot} & Plot coefficient path as a function of $\lambda$ \\
  \bottomrule
\end{tabular}
\label{tab3:list}
\caption{Currently available functions in \CRANpkg{fastcmprsk} (v.1.1.0).}
\end{table}

\subsection{Simulating competing risks data}
Researchers can simulate two-cause competing risks data using the \code{simulateTwoCauseFineGrayModel} function in \CRANpkg{fastcmprsk}. The data generation scheme follows a similar design to that of \cite{fine1999proportional} and \cite{fu2017penalized}. Given a design matrix $\mathbf{Z} = (\mathbf{z}_1^\prime, \ldots, \mathbf{z}_n^\prime)$, $\bbeta_1$, and $\bbeta_2$, let the cumulative incidence function for cause 1 (the event of interest) be defined as
$F_1(t; \mathbf{z}_i) = \Pr(T_i \leq t, \epsilon_i = 1|\mathbf{z}_i) = 1 - [1 - \pi\{1-\exp(-t)\}]^{\exp(\mathbf{z}_i^\prime\bbeta_1)}$,
which is a unit exponential mixture with mass $1 - \pi$ at $\infty$ when $\mathbf{z}_i = \mathbf{0}$ and where $\pi$ controls the cause 1 event rate. The cumulative incidence function for cause 2 is obtained by setting 
$\Pr(\epsilon_i = 2 | \mathbf{z}_i) = 1 - \Pr(\epsilon_i = 1|\mathbf{z}_i)$ and then using an exponential distribution with rate $\exp(\mathbf{z}^\prime_i\bbeta_2)$ for the conditional cumulative incidence function $\Pr(T_i \leq t|\epsilon_i = 2, \mathbf{z}_i)$. Censoring times are independently generated from a uniform distribution $U(u\mbox{\tiny{min}}, u\mbox{\tiny{max}})$ where $u\mbox{\tiny{min}}$ and $u\mbox{\tiny{max}}$ control the censoring percentage. Appendix \ref{app3:data} provides more details on the data generation process.  Below is a toy example of simulating competing risks data where $n = 500$, $\bbeta_1  = (0.40, -0.40, 0, -0.50, 0, 0.60, 0.75, 0, 0, -0.80)$, $\bbeta_2  =-\bbeta_1$, $u\mbox{\tiny{min}} = 0$, $u\mbox{\tiny{max}} = 1$, $\pi = 0.5$, and where $\mathbf{Z}$ is simulated from a multivariate standard normal distribution with unit variance. This simulated dataset will be used to illustrate the use of the different modeling functions within \CRANpkg{fastcmprsk}. { The purpose of the simulated dataset is to demonstrate the use of the \CRANpkg{fastcmprsk} package and its comparative estimation performance to currently-used packages for unpenalized and penalized Fine-Gray regression. Runtime comparisons between the different packages are reported in Section \ref{s3:sim}}.

{
\begin{example}
R> library(fastcmprsk)
R> set.seed(2019)
R> nobs  <- 500 # Set number of observations

R> # Create coefficient vector for event of interest and competing event
R> beta1 <- c(0.40, -0.40,  0, -0.50,  0,  0.60,  0.75,  0,  0, -0.80)
R> beta2 <- -beta1

R> # Simulate design matrix
R> Z     <- matrix(rnorm(nobs * length(beta1)), nrow = nobs)

R> # Generate data
R> dat   <- simulateTwoCauseFineGrayModel(nobs, beta1, beta2, 
+	Z, u.min = 0, u.max = 1, p = 0.5)

R> # Event counts (0 = censored; 1 = event of interest; 2 = competing event)
R> table(dat$fstatus) 

  0   1   2 
241 118 141 

R> # First 6 observed survival times
R> head(dat$ftime)

[1] 0.098345608 0.008722629 0.208321175 0.017656904 0.495185038 0.222799124
\end{example}
}

\subsection{fastCrr: Unpenalized parameter estimation and inference}
We first illustrate the coefficient estimation from (\ref{eq3:pshmodel}) using the Fine-Gray log-pseudo likelihood. 
The \code{fastCrr} function returns a \code{fcrr} object that estimates these parameters using our forward-backward scan algorithm and is syntactically similar to the \code{coxph} function in \CRANpkg{survival}. The \code{formula} argument requires a newly-defined \code{Crisk} object as an outcome. The \code{Crisk} function produces a \code{Crisk} object by calling the \code{Surv} function in \CRANpkg{survival}, modifying it to allow for more than one event, and requires four arguments: a vector of observed event times (\code{ftime}), a vector of corresponding event/censoring indicators (\code{fstatus}), the value of \code{fstatus} that denotes a right-censored observation (\code{cencode}) and the value of \code{fstatus} that denotes the event of interest (\code{failcode}). By default, \code{Crisk} assumes that \code{cencode = 0} and \code{failcode = 1}. { The \code{variance} passed into \code{fastCrr} specifies whether or not the variance should be calculated with parameter estimation.} 
\begin{example}
# cmprsk package
R> fit1 <- crr(dat$ftime, dat$fstatus, Z, failcode = 1, cencode = 0,
+                      variance = FALSE)

# fastcmprsk package
R> fit2 <- fastCrr(Crisk(dat$ftime, dat$fstatus, cencode = 0, failcode = 1) ~ Z,
+                           variance = FALSE)

R> max(abs(fit1$coef - fit2$coef)) # Compare the coefficient estimates for both methods

[1] 8.534242e-08
\end{example}
As expected, the \code{fastCrr} function calculates nearly identical parameter estimates to the \code{crr} function. {The slight difference in numerical accuracy can be explained by the different methods of optimization and convergence thresholds used for parameter estimation. Convergence within the cyclic coordinate descent algorithm used in \code{fastCrr} is determined by the relative change of the coefficient estimates. We allow users to modify the maximum relative change and maximum number of iterations used for optimization within \code{fastCrr} through the \code{eps} and \code{iter} arguments, respectively. By default, we set \code{eps = 1E-6} and \code{iter = 1000} in both our unpenalized and penalized optimization methods.}

We now show how to obtain the variance-covariance matrix for the parameter estimates. The variance-covariance matrix for $\hat{\bbeta}$ { via (\ref{eq3:asymptotic})} can not be directly estimated using the \code{fastCrr} function. First, the asymptotic expression requires estimating both $\Omega$ and $\Sigma$, which can not be trivially calculated in { $O(pn)$ operations}. Second, for large-scale data where both $n$ and $p$ can be large, matrix calculations, storage, and inversion can be computationally prohibitive. Instead, we propose to estimate the variance-covariance matrix using the bootstrap \citep{efron1979bootstrap}.  Let $\tilde{\bbeta}^{(1)}, \ldots \tilde{\bbeta}^{(B)}$ be bootstrapped parameter estimates obtained by resampling subjects with replacement from the original data $B$ times. Unless otherwise noted, the size of each resample is the same as the original data. For $j = 1, \ldots, p$ and $k = 1, \ldots, p$, we can estimate the covariance between $\hat{\beta}_j$ and $\hat{\beta}_k$ by
 \begin{align}
\widehat{Cov}(\hat{\beta}_j, \hat{\beta}_k) = \frac{1}{B - 1} \sum_{b = 1}^B (\tilde{\beta}^{(b)}_{j} - \bar{\beta}_j)(\tilde{\beta}^{(b)}_{k} - \bar{\beta}_k),
\end{align}
where $\bar{\bbeta}_j = \frac{1}{B} \sum_{b=1}^B \tilde{\bbeta}^{(b)}_{j}$. Therefore, with $\hat{\sigma}^2_j = \widehat{Cov}(\hat{\beta}_j, \hat{\beta}_j$), a $(1 - \alpha) \times 100\%$ confidence interval for $\bbeta_j$ is given by
\begin{align}
\hat{\beta}_j \pm z_{1-\alpha/2} \hat{\sigma}_j,
\end{align}
where $z_{1 - \alpha / 2}$ is the $(1 - \alpha) \times 100th$ percentile of the standard normal distribution. Since parameter estimation for the Fine-Gray model can be done in linear time using our forward-backward scan algorithm, the collection of parameter estimates obtained by bootstrapping can also be obtained linearly. The \code{varianceControl} function controls the parameters used for bootstrapping, that one then passes into the \code{var.control} argument in \code{fastCrr}. {These arguments include \code{B}, the number of bootstrap samples to be used and \code{seed}, a non-negative numeric integer to set the seed for resampling}. 

{
\begin{example}
R> # Estimate variance via 100 bootstrap samples using seed 2019.
R> vc   <- varianceControl(B = 100, seed = 2019)
R> fit3 <- fastcmprsk::fastCrr(Crisk(dat$ftime, dat$fstatus) ~ Z, variance = TRUE,
+                              var.control = vc,
+                              returnDataFrame = TRUE)
# returnDataFrame = TRUE is necessary for CIF estimation (next section)

R> round(sqrt(diag(fit3$var)), 3) # Standard error estimates rounded to 3rd decimal place

[1] 0.108 0.123 0.085 0.104 0.106 0.126 0.097 0.097 0.104 0.129
\end{example}
}

{ The accuracy of the bootstrap variance-covariance matrix compared to the asymptotic expression depends on several factors including the sample size and number of bootstrap samples $B$. Our empirical evidence in Section \ref{s3:crr} show that $B = 100$ bootstrap samples provided a sufficient estimate of the variance-covariance matrix for large enough $n$ in our scenarios. In practice, we urge users to increase the number of bootstrap samples, until the variance is stable, if they can computationally afford to. While this may hinder the computational performance of \code{fastCrr} for small sample sizes, we find this to be a more efficient approach for large-scale competing risks data.}

{ We adopt several S3 methods that work seamlessly with the \code{fcrr} object that is outputted from \code{fastCrr}. The \code{coef} method returns the estimated regression coefficient estimates $\hat{\bbeta}$:
\begin{example}
R> coef(fit3) # Coefficient estimates

[1] 0.192275755 -0.386400287  0.018161906 -0.397687129  0.105709092  0.574938015  
[7] 0.778842652 -0.006105756 -0.065707434 -0.996867883
\end{example}

The model pseudo log-likelihood can also be extracted via the \code{logLik} function:
\begin{example}
R> logLik(fit3) # Model log-pseudo likelihood
[1] -590.3842
\end{example}

Related quantities to the log-pseudo likelihood are information criteria, measures of the quality of a statistical model that are used to compare alternative models on the same data. These criterion are computed using the following formula: $-2 l(\hat{\bbeta}) + k \times |\hat{\bbeta}|_0$, where $k$ is a penalty factor for model complexity and $|\hat{\bbeta}|_0$ corresponds to the number of parameters in the model. Information criteria can be computed for a \code{fcrr} object using \code{AIC} and users specify the penalty factor using the \code{k} argument. By default \code{k = 2} and corresponds to the Akaike information criteria \citep{akaike1974new}.  
\begin{example}
R> AIC(fit3, k = 2) # Akaike's Information Criterion
[1] 1200.768

R> # Alternative expression of the AIC
R> -2 * logLik(fit3) + 2 * length(coef(fit3))
[1] 1200.768
\end{example}

If the \code{variance} is set to \code{TRUE} for the \code{fastCrr} model fit, we can extract the bootstrap variance-covariance matrix using \code{vcov}. Additionally, \code{conf.int} will display confidence intervals, on the scale of $\hat{\bbeta}$, and the \code{level} argument can be used to specify the confidence level. By default \code{level = 0.95} and corresponds to $95\%$ confidence limits. 
\begin{example}
R> vcov(fit3)[1:3, 1:3] # Variance-covariance matrix for the first three estimates

             [,1]         [,2]         [,3]
[1,] 0.0116785745 0.0031154634 0.0007890851
[2,] 0.0031154634 0.0150597898 0.0004681825
[3,] 0.0007890851 0.0004681825 0.0072888011


R> confint(fit3, level = 0.95) # 95 

           2.5
x1  -0.01953256  0.4040841
x2  -0.62692381 -0.1458768
x3  -0.14916899  0.1854928
x4  -0.60197206 -0.1934022
x5  -0.10199838  0.3134166
x6   0.32827237  0.8216037
x7   0.58798896  0.9696963
x8  -0.19610773  0.1838962
x9  -0.26995659  0.1385417
x10 -1.24897861 -0.7447572
\end{example}

Lastly, \code{summary} will return an ANOVA table for the fitted model. The table presents the log-subdistribution hazard ratio (\code{coef}), the subdistribution hazard ratio (\code{exp(coef)}), the standard error of the log-subdistribution hazards ratio (\code{se(coef)}) if \code{variance = TRUE} in \code{fastCrr}, the corresponding $z$-score (\code{z value}), and two-sided $p$-value (\code{Pr(|z|)}). When setting \code{conf.int = TRUE}, the \code{summary} function will also print out the $95\%$ confidence intervals (if \code{variance = TRUE} when running \code{fastCrr}). Additionally the pseudo log-likelihood for the estimated model and the null pseudo log-likelihood (when $\hat{\bbeta} = \mathbf{0}$) are also reported below the ANOVA table.

\begin{example}
R> # ANOVA table for fastCrr
R> summary(fit3, conf.int = TRUE) # conf.int = TRUE allows for 95

Fine-Gray Regression via fastcmprsk package. 

fastCrr converged in 24 iterations.
 
Call:
fastcmprsk::fastCrr(Crisk(dat$ftime, dat$fstatus) ~ Z, variance = TRUE, 
    var.control = vc, returnDataFrame = TRUE)

        coef exp(coef) se(coef) z value Pr(>|z|)
x1   0.19228     1.212   0.1081   1.779  7.5e-02
x2  -0.38640     0.679   0.1227  -3.149  1.6e-03
x3   0.01816     1.018   0.0854   0.213  8.3e-01
x4  -0.39769     0.672   0.1042  -3.816  1.4e-04
x5   0.10571     1.111   0.1060   0.997  3.2e-01
x6   0.57494     1.777   0.1259   4.568  4.9e-06
x7   0.77884     2.179   0.0974   7.998  1.3e-15
x8  -0.00611     0.994   0.0969  -0.063  9.5e-01
x9  -0.06571     0.936   0.1042  -0.631  5.3e-01
x10 -0.99687     0.369   0.1286  -7.750  9.1e-15

    exp(coef) exp(-coef)  2.5
x1      1.212      0.825 0.981 1.498
x2      0.679      1.472 0.534 0.864
x3      1.018      0.982 0.861 1.204
x4      0.672      1.488 0.548 0.824
x5      1.111      0.900 0.903 1.368
x6      1.777      0.563 1.389 2.274
x7      2.179      0.459 1.800 2.637
x8      0.994      1.006 0.822 1.202
x9      0.936      1.068 0.763 1.149
x10     0.369      2.710 0.287 0.475
Pseudo Log-likelihood = -590 
Null Pseudo Log-likelihood = -675 
Pseudo likelihood ratio test = 170 on 10 df. 
\end{example}
}

{
Since standard error estimation is performed via bootstrap and resampling, it is easy to use multiple cores to speed up computation. Parallelization is seamlessly implemented using the \CRANpkg{doParallel} package \citep{doParallel}. Enabling usage of multiple cores is done through the \code{useMultipleCores} argument within the \code{varianceControl} function. To avoid interference with other processes, we allow users to set up the cluster on their own. We provide an example below.

\begin{example}
R> library(doParallel) 

R> n.cores <- 2 # No. of cores
R> myClust <- makeCluster(n.cores)

R> # Set useMultipleCores = TRUE to enable parallelization
R> vc = varianceControl(B = 1000, useMultipleCores = TRUE)

R> registerDoParallel(myClust)
R> fit3 <- fastCrr(Crisk(dat$ftime, dat$fstatus) ~ Z, variance = TRUE,
+                             var.control = vc)
R> stopCluster(myClust)
\end{example}
}

\subsection{Cumulative incidence function and interval/band estimation}
The CIF is also available in linear time in the \CRANpkg{fastcmprsk} package. \cite{fine1999proportional} propose a Monte Carlo simulation method for interval and band estimation. We implement a slightly different approach using bootstrapping for interval and band estimation in our package. Let $\tilde{F}^{(1)}_1(t; \mathbf{z}_0), \ldots, \tilde{F}^{(B)}_1(t; \mathbf{z}_0)$ be the bootstrapped predicted CIF obtained by resampling subjects with replacement from the original data $B$ times and let $m(\cdot)$ be a known, monotone, and continuous transformation. In our current implementation we let $m(x) = \log\{-\log(x)\}$; however, we plan on incorporating other transformations in our future implementation. We first estimate the variance function $\sigma^2(t; \mathbf{z}_0)$ of the transformed CIF through
\begin{align}
\label{eq3:boot_var_est}
\hat{\sigma}^2(t; \mathbf{z}_0) = \frac{1}{B} \sum_{b=1}^B \left[ m\{\tilde{F}^{(b)}_1(t; \mathbf{z}_0)\} - \bar{m}\{\tilde{F}_1(t; \mathbf{z}_0)\} \right]^2,
\end{align}
where $\bar{m}\{\tilde{F}_1(t; \mathbf{z}_0)\}  = \frac{1}{B} \sum_{b=1}^B m\{\tilde{F}^{(b)}_1(t; \mathbf{z}_0)\}$. Using the functional delta method, we can now construct $(1 - \alpha) \times 100\%$ confidence intervals for $F_1(t; \mathbf{z}_0)$ by 
\begin{align}
\label{eq3:boot_cif_int}
m^{-1} \left[ m\{\hat{F}_1(t; \mathbf{z}_0)\} \pm z_{1 - \alpha / 2} \hat{\sigma}(t; \mathbf{z}_0)\right].
\end{align}

Next we propose a symmetric global confidence band for the estimated CIF $\hat{F}_1(t; \mathbf{z}_0)$, $t \in [t_L, t_U]$ via bootstrap. We first determine a critical region $C_{1 - \alpha}(\mathbf{z}_0)$ such that 
\begin{align}
\Pr \left\{ \sup_{t \in [t_L, t_U]}  \frac{| m\{\hat{F}_1(t; \mathbf{z}_0)\} - m\{F_1(t; \mathbf{z}_0)\} |}{\sqrt{\widehat{Var}[m\{\hat{F}_1(t; \mathbf{z}_0)\}]}} \leq C_{1 - \alpha}(\mathbf{z}_0) \right\} = 1 - \alpha. 
\end{align}
While Equation (\ref{eq3:boot_var_est}) estimates $\widehat{Var}[m\{\hat{F}_1(t; \mathbf{z}_0)\}]$ we still need to find $C_{1 - \alpha}(\mathbf{z}_0)$ by the bootstrap $(1-\alpha)^{th}$ percentile of the distribution of the supremum in the equation above.  The algorithm is as follows:
\begin{enumerate}
\item Resample subjects with replacement from the original data $B$ times and estimate $\tilde{F}^{(b)}_1(t; \mathbf{z}_0)$ for $b = 1, \ldots, B$ and $\hat{\sigma}^2(t; \mathbf{z}_0)$ using (\ref{eq3:boot_var_est}).
\item For the $b^{th}$ bootstrap sample , $b \in \{1, \ldots, B\}$, calculate
\begin{align*}
C^{(b)} = \sup_{t \in [t_L, t_U]}  \frac{| m\{\tilde{F}^{(b)}_1(t; \mathbf{z}_0)\} - m\{\hat{F}_1(t; \mathbf{z}_0)\} |}{\hat{\sigma}(t; \mathbf{z}_0)}. 
\end{align*}
\item Estimate $C_{1 - \alpha}(\mathbf{z}_0)$ from the sample $(1 - \alpha)^{th}$ percentile of the $B$ values of $C^{(b)}$, denoted by $\hat{C}_{1 - \alpha}(\mathbf{z}_0)$.
\end{enumerate}

Finally, the $(1 - \alpha) \times 100\%$ confidence band for $F_{1}(t; \mathbf{z}_0)$, $t \in [t_L, t_U]$ is given by 
\begin{align}
\label{eq3:boot_cif_band}
m^{-1} \left[ m\{\hat{F}_1(t; \mathbf{z}_0)\} \pm \hat{C}_{1 - \alpha }(\mathbf{z}_0) \hat{\sigma}(t; \mathbf{z}_0)\right].
\end{align}

\begin{figure}[t!]
\label{fig2:cif}
\centering
\includegraphics[scale = 0.6]{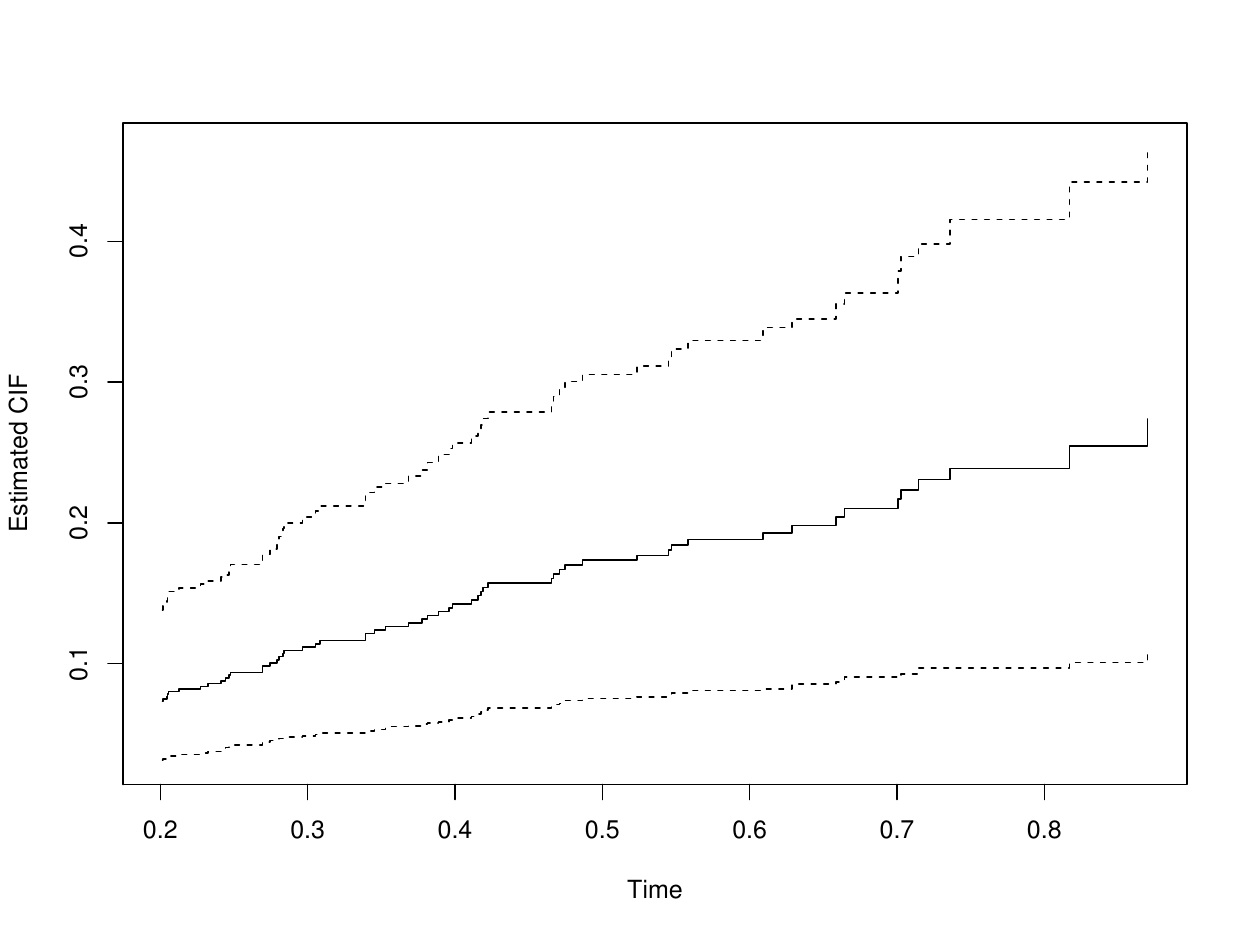}
\caption{Estimated CIF (solid line) and corresponding $95\%$ confidence intervals (dotted lines) between $t_L = 0.2$ and $t_U = 0.9$ given a covariate vector \code{z0} using the coefficient and baseline estimates from our toy example.}
\end{figure}

{ Similar to estimating the variance-covariance matrix for the coefficient estimates $\hat{\bbeta}$, specifying the number of bootstrap samples, seed for reputability, and multicore functionality for estimating the variance of the CIF can be done through the \code{varianceControl} function.} One can perform CIF estimation and interval/band estimation using the \code{predict} function { by specifying a vector $\mathbf{z}_0$ in the \code{newdata} argument and the fitted model from \code{fastCrr}. To calculate the CIF, both the Breslow estimator of the cumulative subdistribution hazard and the (ordered) model data frame need to be returned values within the fitted object. This can be achieved by setting both the \code{getBreslowJumps} and \code{returnDataFrame} arguments within \code{fastCrr} to \code{TRUE}. Additionally, for confidence band estimation one must specify a time interval $[t_L, t_U]$. The user can specify the interval range using the \code{tL} and \code{tU} arguments in \code{predict}. Figure 1 illustrates the estimated CIF and corresponding $95\%$ confidence interval, obtained using 100 bootstrap samples, over the range $[0.2, 0.9]$ given covariate entries \code{z0} simulated from a standard random normal distribution.}

\begin{example}
R> set.seed(2019)
R> # Make sure getBreslowJumps and returnDataFrame are set to TRUE
R> fit4 <- fastCrr(Crisk(dat$ftime, dat$fstatus, cencode = 0, failcode = 1) ~ Z,
+                           variance = FALSE,
+                           getBreslowJumps = TRUE, # Default = TRUE
+                           returnDataFrame = TRUE) # Default is FALSE for storage purposes
 
R> z0 <- rnorm(10) # New covariate entries to predict
R> cif.point <- predict(fit4, newdata = z0, getBootstrapVariance = TRUE,
+                      type = "interval", tL = 0.2, tU = 0.9,
+                      var.control = varianceControl(B = 100, seed = 2019))

R> plot(cif.point) # Figure 1 (Plot of CIF and 95
\end{example}

\subsection{fastCrrp: Penalized Fine-Gray regression in linear time}
\label{s3:fastcrrp}
We extend our forward-backward scan approach for for penalized Fine-Gray regression as described in Section \ref{s3:pen}.  The \code{fastCrrp} function performs LASSO, SCAD, MCP, and ridge \citep{hoerl1970ridge} penalization. {Users specify the penalization technique through the \code{penalty} argument.} The advantage of implementing this algorithm for penalized Fine-Gray regression is two fold. Since the cyclic coordinate descent algorithm used in the \code{crrp} function calculates the gradient and Hessian diagonals in $O(pn^2)$ time, as opposed to $O(pn)$ using our approach, we expect to see drastic differences in runtime for large sample sizes. Second, as mentioned earlier, researchers generally tune the strength of regularization through multiple model fits over a grid of candidate tuning parameter values. Thus the difference in runtime between both methods grows larger as the number of candidate values increases. Below we provide an example of performing LASSO-penalized Fine-Gray regression using {a prespecified grid of 25 candidate values for $\lambda$ that we input into the \code{lambda} argument of \code{fastCrrp}. If left untouched (i.e. \code{lambda = NULL}), a log-spaced interval of $\lambda$ will be computed such that the largest value of $\lambda$ will correspond to a null model. Figure 2 illustrates the solution path for the LASSO-penalized regression, a utility not directly implemented within the \code{crrp} package}. The syntax for \code{fastCrrp} is nearly identical to the syntax for \code{crrp}. 
\begin{example}
R> library(crrp)
R> lam.path <- 10^seq(log10(0.1), log10(0.001), length = 25)

R> # crrp package
R> fit.crrp <- crrp(dat$ftime, dat$fstatus, Z, penalty = "LASSO",
+                         lambda = lam.path, eps = 1E-6)

R> # fastcmprsk package
R> fit.fcrrp <- fastCrrp(Crisk(dat$ftime, dat$fstatus) ~ Z, penalty = "LASSO",
+                                    lambda = lam.path)


R> # Check to see the two methods produce the same estimates.
R> max(abs(fit.fcrrp$coef - fit.crrp$beta))

[1] 1.110223e-15

R> plot(fit.fcrrp) # Figure 2 (Solution path)
\end{example}

\begin{figure}[t!]
\label{fig2:lasso}
\centering
\includegraphics[scale = 0.6]{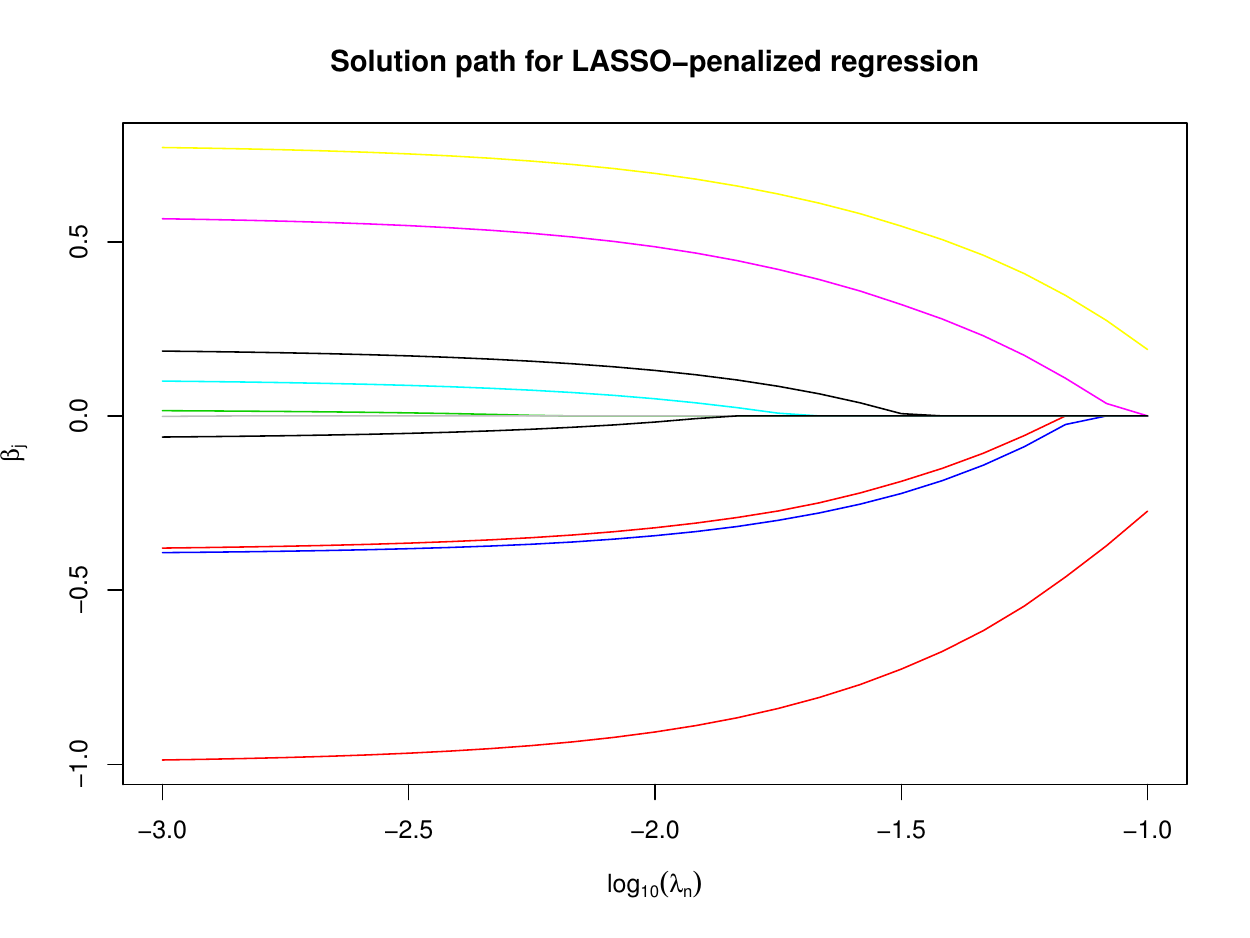}
\caption{Path plot for LASSO-penalized Fine-Gray regression using our toy example. The tuning parameter $\lambda$ varies between the log-spaced interval $[0.001, 0.1]$. The $y$-axis corresponds to the estimated value for $\hat{\beta}_j$ and the $x$-axis corresponds to $\lambda$ (on the $\log_{10}$ scale).}
\end{figure}

\section{Simulation studies}
\label{s3:sim}
This section provides a more comprehensive illustration of the computational performance of the \CRANpkg{fastcmprsk} package over two popular competing packages \CRANpkg{cmprsk} and \CRANpkg{crrp}. 
We simulate datasets under various sample sizes and fix the number of covariates $p = 100$. We generate the design matrix, $\mathbf{Z}$
from a $p$-dimensional standard normal distribution with mean zero, unit variance, and pairwise correlation $\mbox{corr}(z_i, z_j) = \rho^{|i-j|}$, where $\rho = 0.5$ simulates moderate correlation. For Section \ref{s3:crr}, the vector of regression parameters for cause 1, the cause of interest, is $\bbeta_1 = (\bbeta^*, \bbeta^*, \ldots, \bbeta^*)$, where  $\bbeta^* = (0.40, -0.40, 0, -0.50, 0, 0.60, 0.75, 0, 0, -0.80)$. For Section \ref{s3:crrp}, $\bbeta_1 = (\bbeta^*, \mathbf{0}_{p - 10})$. We let $\bbeta_2 = -\bbeta_1$. We set $\pi =  0.5$, which corresponds to a cause 1 event rate of approximately $41\%$.  The average censoring percentage for our simulations varies between $30-35\%$.  We use  \code{simulateTwoCauseFineGrayModel} to simulate these data and average { runtime} results over 100 Monte Carlo replicates. We report timing on a system with an Intel Core i5 2.9 GHz processor and 16GB of memory.

\subsection[Comparison to the crr package]{Comparison to the \CRANpkg{crr} package}
\label{s3:crr}
In this section, we compare the runtime and estimation performance of the \code{fastCrr} function to \code{crr}. We vary $n$ from $1,000$ to { 500,000} 
and run \code{fastCrr} and \code{crr} both with and without variance estimation. We take 100 bootstrap samples, {without parallelization}, to obtain the bootstrap standard errors with \code{fastCrr}. { As shown later in the section (Tables 3 and 4), 100 bootstrap samples suffices to produce a good standard error estimate with close-to-nominal coverage for large enough sample sizes 
in our scenarios. In practice, we recommend users to increase the number of bootstrap samples until the variance estimate becomes stable, when computationally feasible.}

{ Figure 3 depicts how fast the computational complexities of   \code{fastCrr} (dashed lines) and \code{crr} (solid lines) increase as $n$ increases as measured by runtime  (in seconds)}. It shows clearly that the computational complexity of \code{crr}  increases quadratically ({ solid line slopes $\approx 2$}) while that of \code{fastCrr} is linear ({ dashed line slopes $\approx 1$}). { This implies that  the computational gains of \code{fastCrr} over \code{crr}  are expected to grow exponentially as the sample size increases.  
\begin{figure}[t]
\centering
\includegraphics[scale = 0.6]{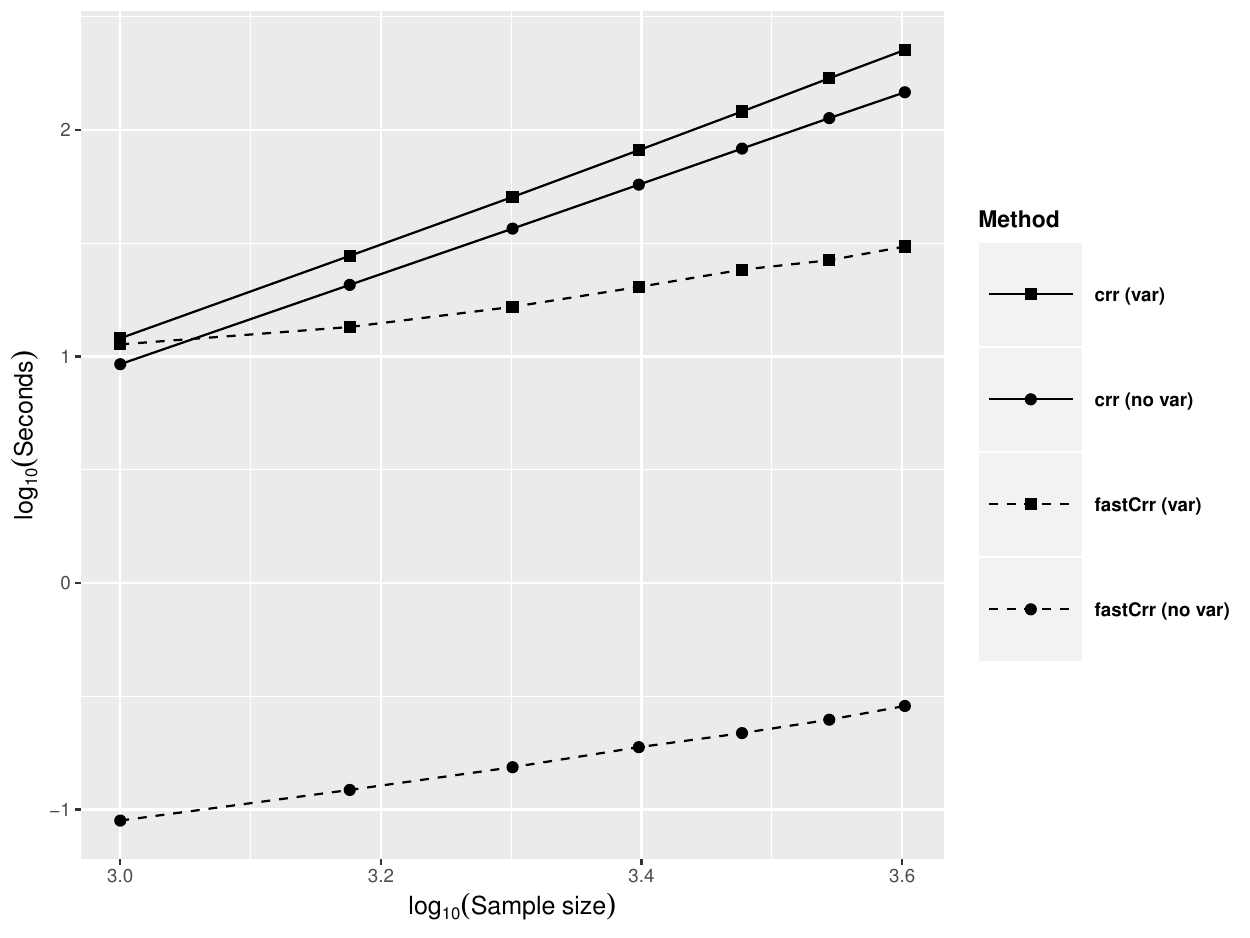}
\label{fig3:timing}
\caption{Runtime comparison between \code{fastCrr} and \code{crr} with and without variance estimation. Axes are on the $\log_{10}$ scale. Solid and dashed lines represent the \CRANpkg{crrp} and \CRANpkg{fastcmprsk} implementation, respectively. Square, and circle symbols denote variance and without variance calculation, respectively. Variance estimation for \code{crr} is performed using the asymptotic expression of the variance-covariance estimator. Variance estimation for \code{fastCrr} is performed using 100 bootstrap samples. Reported runtime are averaged over 100 Monte Carlo runs.}
\end{figure}

We further demonstrates the computational advantages of \code{fastCrr} over \code{crr} for large sample size data in Table \ref{runtimelarge-n} by comparing their runtime on a single simulated data with  $n$ varying from $50,000$ to $500,000$ using a system with an Intel Xeon 2.40GHz processor and 256GB of memory. It is seen that \code{fastCrr} scales well to  large sample size data, whereas \code{crr} eventually grinds to a halt as $n$ grows large. For example, for $n=500,000$, it only takes less than 1 minute for
\code{fastCrr} to finish, while \code{crr} did not finish in 3 days.
Because the forward-backward scan allows us to efficiently compute variance estimates through bootstrapping, we have also observed drastic computational gains in variance estimation with large sample size data (7 minutes for \code{fastCrr} versus 54 hours for \code{crr}.  Furthermore, since parallelization of the bootstrap procedure was not implemented in these timing reports, we expect multicore usage to further decrease the runtime of the variance estimation for \code{fastCrr}
}
\begin{table}[t]
\centering
\setlength{\tabcolsep}{3.2pt}
\caption{Runtime comparison of \code{crr} versus \code{fasrCrr} for large $n$ scenarios. The dashes (``--") indicate that runtime could not be completed within 72 hours. Variance estimation for \code{fastCrr} is calculated using $B = 100$ bootstrap samples.}
\begin{tabular}{l|lll}
\toprule
&\multicolumn{3}{c}{Sample size $n$}\\ 
  &  50,000  & 100,000  & 500,000  \\ 
  \midrule
    \code{crr} without variance & 6 hours  & 24 hours & --  \\ 
    \code{crr}  with variance & 54 hours  & --  & -- \\
 \code{fastCrr} without variance & 5 seconds  & 12 seconds & 50 seconds  \\ 
 \code{fastCrr} with variance & 7 minutes  & 14 minutes &  69 minutes \\
   \bottomrule
\end{tabular}
\label{runtimelarge-n}
\end{table}


{ We also performed a simulation  to compare the bootstrap procedure for variance estimation to the estimate of the asymptotic variance provided in (\ref{eq3:asymptotic}) used in \code{crr}. First, we compare the two standard error estimates with the empirical standard error of $\hat{\bbeta}_{1}$. For the $j^{th}$ coefficient, the empirical standard error is calculated as the standard deviation of $\hat{\beta}_{1j}$ from the 100 Monte Carlo runs. For the standard error provided by both the bootstrap and the asymptotic variance-covariance matrix, we take the average standard error of $\hat{\beta}_{1j}$ over the 100 Monte Carlo runs. Table \ref{table3} compares the standard errors for $\hat{\beta}_{1j}$ for $j = 1, 2, 3$. When $n = 1000$, the average standard error using the bootstrap is slightly larger than the empirical standard error; whereas, the standard error from the asymptotic expression is slightly smaller. These differences diminish and all three estimates are comparable when $n \geq 2000$. This provides  evidence that both the bootstrap and asymptotic expression are adequate estimators of the variance-covariance expression for large datasets. 

\begin{table}[t]
\centering
\setlength{\tabcolsep}{3.2pt}
\caption{Standard error estimates for various different values of $\beta_{1j}$ ($j = 1, 2, 3$). Empirical: Standard deviation of the 100 Monte Carlo estimates of $\hat{\beta}_{1j}$; Bootstrap: The average of the 100 Monte Carlo estimates of the bootstrap standard error for $\hat{\beta}_{1j}$ using $B = 100$ bootstrap samples; Asymptotic: The average of 100 Monte Carlo estimates of the standard error estimate for $\hat{\beta}_{1j}$ using the asymptotic variance-covariance matrix defined in (\ref{eq3:asymptotic}). }
\begin{tabular}{llrrrr}
  \toprule
 & Std. Err.  Est. & $n=$ 1000 & 2000 & 3000 & 4000 \\ 
  \midrule
$\beta_{11} = 0.4$ & Empirical & 0.06 & 0.05 & 0.04 & 0.03 \\ 
   & Bootstrap & 0.10 & 0.05 & 0.04 & 0.03 \\ 
   & Asymptotic & 0.07 & 0.04 & 0.03 & 0.03 \\ 
   \midrule
$\beta_{12} = -0.4$ & Empirical & 0.10 & 0.05 & 0.04 & 0.03 \\ 
  & Bootstrap & 0.11 & 0.06 & 0.04 & 0.04 \\ 
  & Asymptotic & 0.08 & 0.05 & 0.04 & 0.03 \\ 
  \midrule
$\beta_{13} = 0$ & Empirical & 0.09 & 0.06 & 0.04 & 0.03 \\ 
  & Bootstrap & 0.11 & 0.06 & 0.04 & 0.04 \\ 
  & Asymptotic & 0.07 & 0.05 & 0.04 & 0.03 \\
   \bottomrule
\end{tabular}
\label{table3}
\end{table}

Additionally, we present in Table \ref{tab3:covprob}  the coverage probability (and standard errors) of the $95\%$ confidence intervals for $\beta_{11} = 0.4$ using the bootstrap (\code{fastCrr}) and asymptotic (\code{crr}) variance estimate. The confidence intervals are wider for the bootstrap approach when compared to confidence intervals produced using the asymptotic variance estimator, especially when $n = 1000$. However, both methods are close to the nominal $95\%$ level as $n$ increases. We observe similar trends across the other coefficient estimates. 

\begin{table}[t]
\centering
\setlength{\tabcolsep}{3.2pt}
\caption{Coverage probability (and standard errors) of $95\%$ confidence intervals for $\beta_{11} = 0.4$. Confidence intervals for \code{crr} are calculated using the asymptotic expression of the variance-covariance estimator. Confidence intervals for \code{fasrCrr} are calculated using the bootstrap variance-covariance estimator using 100 bootstrap samples.}
\begin{tabular}{l|rrrr}
\toprule
  & $n$ = 1000  & 2000  & 3000 & 4000 \\ 
  \midrule
    \code{crr} & 0.93 (0.03) & 0.90 (0.03) & 0.93 (0.03) & 0.95 (0.02) \\ 
 \code{fastCrr} & 1.00 (0.00) & 0.98 (0.02) & 0.95 (0.02) & 0.95 (0.02) \\ 
   \bottomrule
\end{tabular}
\label{tab3:covprob}
\end{table}

}


\subsection[Comparison to the crrp package]{Comparison to the \CRANpkg{crrp} package}
\label{s3:crrp}
As mentioned in Section \ref{s3:pen}, \cite{fu2017penalized} provide an {R} package \CRANpkg{crrp} for performing penalized Fine-Gray regression using the LASSO, SCAD, and MCP penalties. We compare the runtime between \code{fastCrrp} with the implementation in the \CRANpkg{crrp} package. To level comparisons, we modify the source code in \code{crrp} so that the function only calculates the coefficient estimates and BIC score. We vary $n = 1000, 1500, \ldots, 4000$, fix $p = 100$,  and employ a  25-value grid search for the tuning parameter. Figure 4 illustrates the computational advantage the \code{fastCrrp}
function has over \code{crrp}. 

{Similar to the unpenalized scenario}, the computational performance of \code{crrp} (solid lines) increases quadratically while \code{fasrCrrp} (dashed lines) increases linearly, resulting in a 200 to 300-fold speed up in runtime when $n = 4000$. {This, along with the previous section and a real data analysis conclusion in the following section, strongly suggests that for large-scale competing risks datasets (e.g. EHR databases), where the sample size can easily exceed tens to hundreds of thousands, analyses that may take several hours or days to perform using currently-implemented methods are available within seconds or minutes using our forward-backward scan algorithm.} 

\begin{figure}[t!]
\centering
\includegraphics[scale = 0.6]{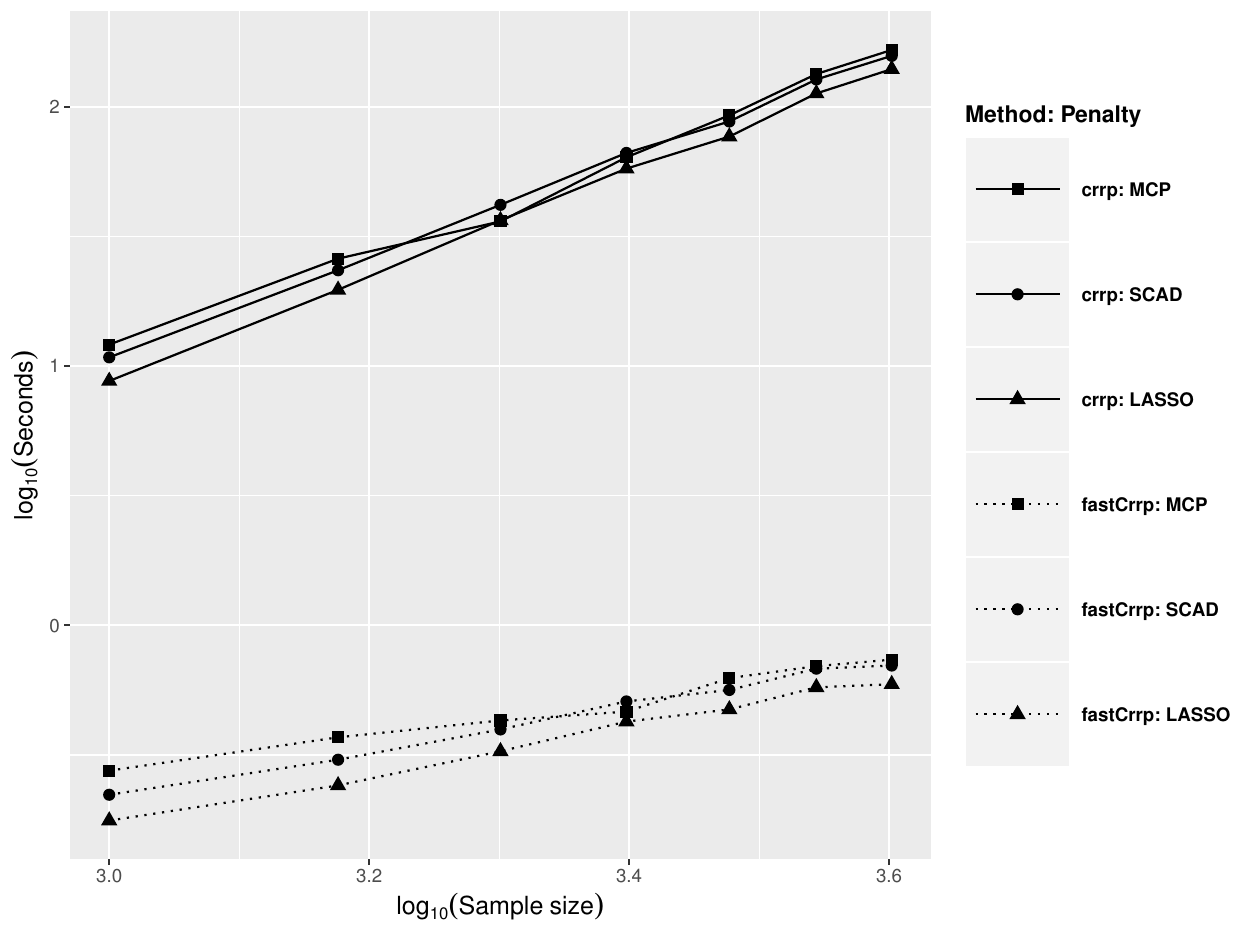}
\label{fig3:pen}
\caption{Runtime comparison between the \CRANpkg{crrp} and \CRANpkg{fastcmprsk} implementations of LASSO, SCAD, and MCP penalization. Solid and dashed lines represent the \CRANpkg{crrp} and \CRANpkg{fastcmprsk} implementation, respectively. Square, circle, and triangle symbols denote the penalties MCP, SCAD, and LASSO, respectively. Axes are on the $\log_{10}$ scale. Reported runtime are averaged over 100 Monte Carlo runs.}
\end{figure}

\section{Discussion}
The \CRANpkg{fastcmprsk} package provides a set of scalable tools for the analysis of large-scale competing risks data by developing an approach to linearize the computational complexity required to estimate the parameters of the Fine-Gray proportional subdistribution hazards model. Multicore use is also implemented to further speed up methods that require bootstrapping and resampling. Our simulation results show that our implementation results in a up to 7200-fold decrease in runtime for large sample size data. { We also note that in a real-world application, \cite{kawaguchi2019scalable} record a drastic decrease in runtime ($\approx$ 24 hours vs. $\approx$ 30 seconds) when comparing the proposed implementation of LASSO, SCAD, and MCP to the methods available in \CRANpkg{crrp} on a subset of the United States Renal Data Systems (USRDS) where $n =125, 000$.}
The package implements both penalized and unpenalized Fine-Gray regression and we can conveniently extend our forward-backward algorithm to other applications such as stratified and clustered Fine-Gray regression. 

Lastly, our current implementation assumes that covariates are densely observed across subjects. This is problematic in the sparse high-dimensional massive sample size (sHDMSS) domain \citep{mittal2013high} where the number of subjects and sparsely-represented covariates easily exceed tens of thousands. These sort of data are typical in large comparative effectiveness and drug safety studies using massive administrative claims and EHR databases and typically contain millions to hundreds of millions of
patient records with tens of thousands patient attributes,
which such settings are particularly useful  for drug safety studies of a rare event such as unexpected adverse events 
\citep{schuemie2018improving} to protect public health. We are currently extending our algorithm to this domain in a sequel paper.

\section{Acknowledgements}
{ We thank the referees and the editor for their helpful comments that improved the presentation of the article}.
Marc A. Suchard's work is partially supported through the National Institutes of Health grant U19 AI 135995. {Jenny I. Shen's work is partly supported through the National Institutes of Health grant K23DK103972.} The research of Gang Li was partly supported by National Institutes of Health Grants P30 CA-16042, UL1TR000124-02, and P50 CA211015.



\newpage

\begin{appendix}

\section{Data generation scheme} 
\label{app3:data}
We describe the data generation process for the \code{simulateTwoCauseFineGrayModel} function. Let $n$, $p$, $\mathbf{Z}_{n \times p}$, $\bbeta_1$, $\bbeta_2$, $u\mbox{\tiny{min}}$, $u\mbox{\tiny{max}}$ and $\pi$ be specified. We first generate independent Bernoulli random variables to simulate the cause indicator $\ep$ for each subject. That is, $\ep_i \sim 1 + Bern\{(1 - \pi)^{\exp(\mathbf{z}_i^\prime \bbeta_1)}\}$ for $i = 1, \ldots, n$. Then, conditional on the cause, event times are simulated from 
\begin{align*}
\Pr(T_i \leq t | \ep_i = 1, \mathbf{z}_i) & = \frac{1 - [1 - \pi\{1 - \exp(-t)\}]^{\exp(\mathbf{z}_i^\prime\bbeta_1)}}{1 - (1 - \pi)^{\exp(\mathbf{z}_i^\prime\bbeta_1)}} \\
\Pr(T_i \leq t | \ep_i = 2, \mathbf{z}_i) & = 1 - \exp\{-t\exp(\mathbf{z}_i^\prime\bbeta_2)\}, \\
\end{align*}
and $C_i \sim U(u\mbox{\tiny{min}}, u\mbox{\tiny{max}})$. Therefore, for $i = 1, \ldots, n$, we can obtain the following quadruplet $\{(X_i, \delta_i, \delta_i \ep_i, \mathbf{z}_i)\}$ where $X_i = \min(T_i, C_i)$,  and $\delta_i = I(X_i \leq C_i)$. Below is an excerpt of the code used in \code{simulateTwoCauseFineGrayModel} to simulate the observed event times, cause and censoring indicators.

\begin{example}
#START CODE
...
...
...
# nobs, Z, p = pi, u.min, u.max, beta1 and beta2 are already defined.
# Simulate cause indicators here using a Bernoulli random variable
c.ind <- 1 + rbinom(nobs, 1, prob = (1 - p)^exp(Z 

ftime <- numeric(nobs)
eta1 <- Z[c.ind == 1, ] 
eta2 <- Z[c.ind == 2, ] 

# Conditional on cause indicators, we simulate the model.
u1 <- runif(length(eta1))
t1 <- -log(1 - (1 - (1 - u1 * (1 - (1 - p)^exp(eta1)))^(1 / exp(eta1))) / p)
t2 <- rexp(length(eta2), rate = exp(eta2))
ci <- runif(nobs, min = u.min, max = u.max) # simulate censoring times

ftime[c.ind == 1] <- t1
ftime[c.ind == 2] <- t2 
ftime <- pmin(ftime, ci) # X = min(T, C)
fstatus <- ifelse(ftime == ci, 0, 1) # 0 if censored, 1 if event 
fstatus <- fstatus * c.ind  # 1 if cause 1, 2 if cause 2   
...
...
...            
\end{example}

\end{appendix}

\bibliography{kawaguchi}

\address{Eric S. Kawaguchi\\
  University of Southern California\\
  Department of Preventive Medicine \\
  2001 N. Soto St. 
  Los Angeles, CA 90032, 
  USA\\
  \email{eric.kawaguchi@med.usc.edu}}

\address{Jenny I. Shen \\
The Lundquist Institute at Harbor-UCLA Medical Center \\
Division of Nephrology and Hypertension   \\
1124 W. Carson St. \\
Torrance, CA  90502,
 USA \\
  \email{jshen@lundquist.org}}

\address{Gang Li \\
University of California, Los Angeles\\
Departments of Biostatistics and Computational Medicine\\
Los Angeles, CA 90095,
USA \\
  \email{vli@ucla.edu}}

\address{Marc A. Suchard\\
University of California, Los Angeles \\
Departments of Biostatistics, Computational Medicine, and Human Genetics \\
Los Angeles, CA 90095,
USA \\
  \email{msuchard@ucla.edu}}
\end{article}

\end{document}